\documentclass[]{spie}  

\usepackage{amsmath,amsfonts,amssymb}
\usepackage{graphicx}
\usepackage[colorlinks=true, allcolors=blue]{hyperref}
\usepackage{dsfont}
\usepackage{pifont}
\newcommand{\cmark}{\ding{51}}%
\newcommand{\xmark}{\ding{55}}%
\usepackage{harveyballs}
\usepackage{multirow}
\usepackage{colortbl}
\usepackage{tikz}
\usetikzlibrary{calc}

\definecolor{RWTHBlau}{RGB}{0,84, 159}
\definecolor{RWTHHellblau}{RGB}{142,186,229}
\definecolor{RWTHPetrol}{RGB}{0,97,101}
\definecolor{RWTHTuerkis}{RGB}{0,152,161}
\definecolor{RWTHGruen}{RGB}{87,171,39}
\definecolor{RWTHMaigruen}{RGB}{189,205,0}
\definecolor{RWTHGelb}{RGB}{255,237, 0}
\definecolor{RWTHOrange}{RGB}{246,168,0}
\definecolor{RWTHMagenta}{RGB}{227, 0, 102}
\definecolor{RWTHRot}{RGB}{204,7,30}
\definecolor{RWTHBordeaux}{RGB}{161,16,53}
\definecolor{RWTHViolett}{RGB}{97,33,88}
\definecolor{RWTHLila}{RGB}{122,111,172}
\definecolor{RWTHCustomGrau}{gray}{0.8}
\definecolor{RWTHCustomDunkelgrau}{gray}{0.4}
\definecolor{RWTHCustomNachtgrau}{gray}{0.2}

\title{Model-Based Interfacing of Large-Scale Metrology Instruments}

\author[a]{Benjamin Montavon}
\author[a]{Martin Peterek}
\author[a]{Robert H.~Schmitt}

\affil[a]{Laboratory for Machine Tools and Production Enginnering (WZL) of RWTH Aachen University, Campus-Boulevard 30, 52074 Aachen, Germany}

\authorinfo{E-Mail Correspondance: \textit{R.Schmitt@wzl.rwth-aachen.de}}

\begin{document} 
\maketitle

\begin{abstract}
   Metrology assisted assembly systems constitute cyber physical production systems relying on in-process sensor data as input to model-based control loops. These range from local, physical control loops, e.g. for robots to closed-loop product lifecycles including quality management. The variety and amount of involved sensors, actors and data sources require a distinct infrastructure to ensure efficient, reliable and secure implementation. Within the paradigm of Internet of Production a reference architecture for such an infrastructure is established by four layers: Raw data (1), provisioning of proprietary systems (2), data aggregation and brokering (3) and decision support (4).
   In modern metrology assisted assembly systems, a virtual reference frame is constituted by one or multiple predominantly optical Large-Scale Metrology instruments, e.g. laser trackers, indoor GPS or multilateration based on ultra wideband communication. An economically efficient implementation of the reference frame can be achieved using cooperative data fusion, both by increasing the operative volume with existing systems and by optimizing the utilization of highly precise and therefor typically cost-intensive instruments. Herewith a harmonization is required as well from a physical perspective as in terms of communication interfaces to the raw data provided by the individual instruments.  The authors propose a model-based approach to obtain a protocol-agnostic interface description, viewing a Large-Scale Metrology instrument as an abstract object oriented system consisting of one or multiple base units and mobile entities. Its object-oriented structure allows a realization of the interface in arbitrary structured communication protocols by adhering to fixed data transformation schemes. This approach is evaluated for MQTT (structured by topics), OPC UA (structured by data model) and HTTP/REST (structured by URLs) as key protocols within the internet of things. Moreover, the transformation between different protocols decouples software requirements of measurement instruments and actors, generally allowing a more efficient integration into cyberphysical production systems.
\end{abstract}


\section{INTRODUCTION}
Large-Scale Metrology, in addition to its traditional applications, has become a backbone of novel manufacturing paradigms \cite{Schmitt.2016, Huettemann.2016,Maropoulos.2014}. At the same time, the resulting cyberphysical production sytems are of growing complexity, accommodating increasing numbers of heterogeneous sensors and demanding interoperability \cite{Brecher.2017}. Reference architectures as presented in section \ref{sec:iop} formalize the data flow of such systems accounting for a large number of heterogeneous entities. An efficient implementation of such manufacturing systems requires apropriate communication interfaces to the Large-Scale Metrology Instruments. Throughout this paper, a protocol- and device-agnostic approach is described by apropiately modeling IoT protocols and instruments.

From the perspective of information technology, a  measurement can be interpreted as self-contained microservice with defined capability and opaque system details, which is required to interoperate with other services \cite{Wolff.2016}. \textsc{Evertz} et al.~deduce three main concerns for their operability: The representation of data over a communication wire, the order and type of individual items in a message, and the physical meaning and scale of these items \cite{Evertz.2015}. Regarding the latter question, an interface model based on a physical view of a Large-Scale Metrology instrument is developed in section \ref{sec:model}. The first two issues are addressed by IoT protocols, e.g.~OPC UA and MQTT. Common protocols are reviewed in section \ref{sec:protocols} to decouple the defined interface model from a specific protocol by establishing an abstraction layer. This allows the realisation of the microservice interface using mature and widely supported standards and thereby to reduce implementation effort and development environment confinements. 

The entire approach is validated for a laser tracker in section \ref{sec:tracker} used in a cloud-based machine tool calibration application.
\section{CYBER-PHYSICAL SYSTEMS AND INTERNET OF PRODUCTION}
\begin{figure}[th]
\begin{center}
    \includegraphics[width=\textwidth]{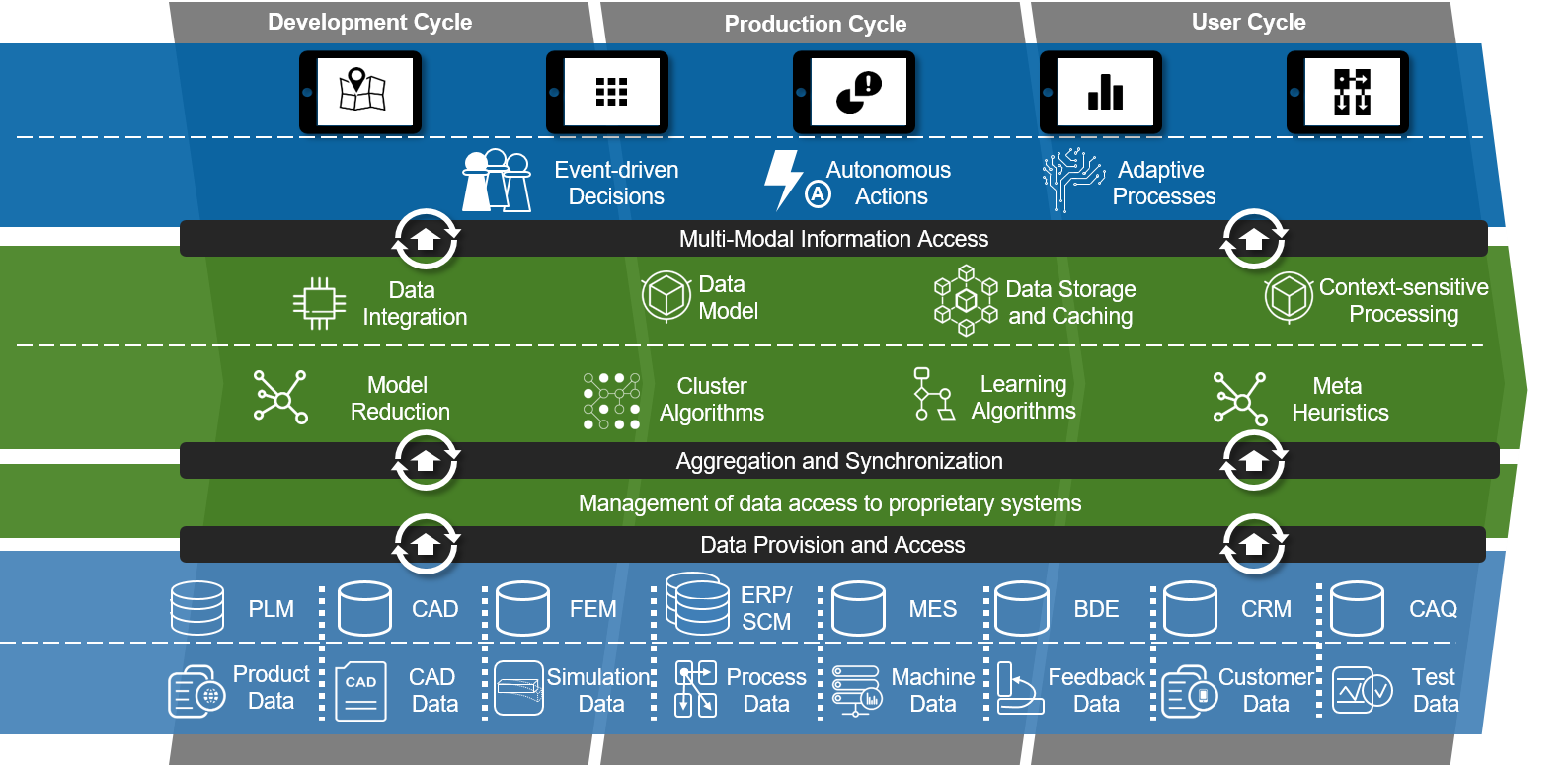}
\end{center}
\caption{Four-layer semantic data architecture for cyberphysical production systems within the concept of Internet of Production \cite{Brecher.2017}.}
\label{fig:iop}
\end{figure}

\label{sec:iop}
Within the concept of the Internet of Production, a semantic four-layer architecture for cyberphysical production systems shown in figure \ref{fig:iop} is proposed \cite{Brecher.2017}. The first layer represents the raw data sources, including sensor data, product and process information. It is followed by a provisioning layer, which is required to enable access to different data sources with heterogeneous communication patterns and protocols. Within the third level, the aggregated data is processed, e.g.~using model-based data reduction or learning algorithms to serve as foundation of data-driven decisions associated with the top layer. The latter is constituted by applications supporting a human user taking a decision and algorithms for automated system control, effectively leading to closed-loop behaviour as the raw data on the lowest level provides feedback signals. The data flow occuring in industrial Large-Scale Metrology can be associated to the aforementioned model: Measurement instruments, nominal models and process descriptions represent the raw data. Geometric analysis and further processing of the measured coordinates reside in the third layer. A digital Geometric Design \& Tolerancing report is an example for user-oriented decision support, while metrology-assisted autonomous vehicle and robot guidance are exemplary autonomous systems within the top layer \cite{Schuppstuhl.2018}.

The model-based interface proposed by the authors addresses the data provisioning layer. With the high variety of existing Large-Scale Metrology instruments and their proprietary interfaces, the integration of a new device into an existing system poses a significant effort \cite{Schmitt.2016, Franceschini.2014}. Even if the communication over the network is harmonized using common IoT protocols, the interface model,i.e.~structure, identifiers and definition of functions and variables can differ and thereby require a manual adaption of existing applications in the processing layer when the instrument providing the data changes. This impacts the selection of a device via its indirect cost, conflicting with the ambition to choose a measurement instrument solely based on the metrologic capabilities required by the application \cite{Montavon.2017}. By structuring the device interface using a generalized functional model prior within the utilized IoT protocols as proposed in section \ref{sec:model}, this limitation can be overcome. 

\section{PROTOCOL-AGNOSTIC INTERFACE IMPLEMENTATION}
\label{sec:protocols}
IoT protocols enable multilateral interfacing among the devices and applications within the defined layers by defining how data is exchanged on an information technical level.
Within most of the protocols, common characteristics can be identified from a practical perspective. In their following descriptions, the term message refers the entirety of data transmitted including protocol-related overhead. 

\begin{table}[p]
   \renewcommand{\arraystretch}{1.25}
   \renewcommand{\arrayrulewidth}{1.25pt}
   \arrayrulecolor{white}
   \caption{Classification of properties and capabilities of for a selection of current IoT protocols from the practical perspective of a developer implementing an interface. \cmark~means native support, while \xmark~relates to undefined within protocol. \harveyBallNone~-~\harveyBallFull~indicate the feasability of a custom implementation partially reusing protocol characteristics. REST is choosen as common paradigm for designing HTTP interfaces.}
   \begin{tabular}{|>{\columncolor{RWTHCustomGrau}}p{0.165\textwidth}|>{\columncolor{RWTHCustomGrau}}p{0.07\textwidth}|>{\columncolor{RWTHCustomGrau}}p{0.07\textwidth}|>{\columncolor{RWTHCustomGrau}}p{0.07\textwidth}|>{\columncolor{RWTHCustomGrau}}p{0.07\textwidth}|>{\columncolor{RWTHCustomGrau}}p{0.07\textwidth}|>{\columncolor{RWTHCustomGrau}}p{0.07\textwidth}|>{\columncolor{RWTHCustomGrau}}p{0.07\textwidth}|>{\columncolor{RWTHCustomGrau}}p{0.07\textwidth}|}
   \hline
   \rowcolor{RWTHHellblau} \cellcolor{white}
   &  \hspace*{3pt}\rotatebox{90}{\textbf{\parbox{0.2\textwidth}{Publish/\\ Subscribe}}}
   &  \hspace*{3pt}\rotatebox{90}{\textbf{\parbox{0.2\textwidth}{Request/\\ Response}}}
   &  \hspace*{3pt}\rotatebox{90}{\textbf{\parbox{0.2\textwidth}{Authentication/ Authorization}}}
   &  \hspace*{3pt}\rotatebox{90}{\textbf{\parbox{0.2\textwidth}{Structured\\ Identifiers}}}
   &  \hspace*{3pt}\rotatebox{90}{\textbf{\parbox{0.2\textwidth}{Serialization\\ Model}}}
   &  \hspace*{3pt}\rotatebox{90}{\textbf{\parbox{0.2\textwidth}{Function\\ Invocation}}}
   &  \hspace*{3pt}\rotatebox{90}{\textbf{\parbox{0.2\textwidth}{CRUD \\ Data View}}}
   &  \hspace*{3pt}\rotatebox{90}{\textbf{\parbox{0.2\textwidth}{Model \\ Browsing}}}
   \\ \hline \textbf{OPC UA} & \hspace*{10pt}\cmark & \hspace*{10pt}\cmark & \hspace*{10pt}\cmark & \hspace*{10pt}\cmark & \hspace*{10pt}\cmark & \hspace*{10pt}\cmark & \hspace*{10pt}\harveyBallHalf & \hspace*{10pt}\cmark
   \\ \hline \textbf{MQTT} & \hspace*{10pt}\cmark & \hspace*{10pt}\xmark & \hspace*{10pt}\cmark & \hspace*{10pt}\cmark & \hspace*{10pt}\xmark & \hspace*{10pt}\harveyBallQuarter & \hspace*{10pt}\harveyBallQuarter & \hspace*{10pt}\xmark
   \\ \hline \textbf{AMQP} & \hspace*{10pt}\cmark & \hspace*{10pt}\xmark & \hspace*{10pt}\cmark & \hspace*{10pt}\cmark & \hspace*{10pt}\xmark & \hspace*{10pt}\harveyBallQuarter & \hspace*{10pt}\harveyBallQuarter & \hspace*{10pt}\xmark
   \\ \hline 
   \textbf{HTTP/REST} & \hspace*{10pt}\xmark &  \hspace*{10pt}\cmark  &  \hspace*{10pt}\harveyBallFull &  \hspace*{10pt}\cmark &  \hspace*{10pt}\harveyBallQuarter &  \hspace*{10pt}\harveyBallThreeQuarter &  \hspace*{10pt}\cmark & \hspace*{10pt}\cmark
   \\ \hline 
   \textbf{GRPC} &  \hspace*{10pt}\harveyBallQuarter &  \hspace*{10pt}\cmark &  \hspace*{10pt}\harveyBallHalf &  \hspace*{10pt}\xmark &  \hspace*{10pt}\cmark &  \hspace*{10pt}\cmark  &  \hspace*{10pt}\xmark & \hspace*{10pt}\xmark
   \\ \hline \textbf{BLE} &  \hspace*{10pt}\cmark    &  \hspace*{10pt}\cmark  &  \hspace*{10pt}\cmark  &  \hspace*{10pt}\xmark &  \hspace*{10pt}\harveyBallHalf &  \hspace*{10pt}\harveyBallNone &  \hspace*{10pt}\harveyBallNone & \hspace*{10pt}\harveyBallHalf \\
   \end{tabular}

   \label{tab:protocol_characteristics}

   \quad
   \vspace*{48pt}
   \quad

   \renewcommand{\arraystretch}{1.25}
   \renewcommand{\arrayrulewidth}{1.25pt}
   \arrayrulecolor{white}
   \caption{Overview of resource actions and their significance to objects (O), variables (V) and functions (F). Depending on the messaging pattern, only a subset of actions is called. ON\_SUBSCRIBE and ON\_UNSUBSCRIBE may be delegated to a protocol-specific message broker. The provided examples are non-exclusive.}
   \begin{tabular}{|>{\columncolor{RWTHCustomGrau}}p{0.26\textwidth}|>{\columncolor{RWTHCustomGrau}}c|>{\columncolor{RWTHCustomGrau}}p{0.34\textwidth}|>{\columncolor{RWTHCustomGrau}}p{0.26\textwidth}|}  \hline
       \rowcolor{RWTHHellblau} \textbf{Action} & & \textbf{Significance} & \textbf{Example(s)} \\ \hline
       READ & \textbf{V} & Retrieve a variable's current value. & HTTP GET, OPC UA read\\ \hline
       UPDATE & \textbf{V} & Set a variable's value. If this is not meaningful (e.g.~for quantities measured by a sensor), the authorization logic should return read-only permissions. & HTTP PUT, BLE write characteristic \\ \hline
       CREATE & \textbf{O} & Update the device/service by adding a resource. & HTTP POST\\ \hline
       DELETE & \textbf{O} & Update the device/service by removing a resource. & HTTP DELETE\\ \hline
       ON\_SUBSCRIBE& \textbf{V} & Add a recipient to value update notifications of the variable. & MQTT subscribe, GRPC stream, OPC UA subscription\\ \hline
       ON\_UNSUBSCRIBE & \textbf{V} & Remove an update recipient. & MQTT unsubscribe\\ \hline
       NOTIFY  & \textbf{V} &  Callback dispatching value changes to recipients. & MQTT publish, GRPC stream, OPC UA publish\\ \hline
       ON\_INVOKE & \textbf{F} & Invoke an implemented function. & OPC UA method, GRPC call\\ \hline
   \end{tabular}
   \label{tab:actions}
\end{table}

\noindent\textbf{Messaging Pattern} \\
In a \textit{Request/Response} pattern, the exchange is initiated by the client sending a request to a server responding according to message's content. Multiple clients typically use independent channels. With a \textit{Publish/Subscribe} pattern, channels are shared, such that the exchange is initiated by the party sending or listening to a specific message channel, often referred to as topics. A protocol can support both patterns.

\noindent\textbf{Serialization Model} \\
The minimum requirement to employ a communication protocol is a defined extraction of raw, non protocol-related data from the message. If a serialization scheme can be defined within the protocol, the raw data is directly decoded into typed data fields. Otherwise, a serialized representation can be established using general schemes such as JSON, XML or YAML.

\noindent\textbf{Authentication and Authorization} \\
Authentication is the functionality of a protocol to identify the entities participating in the message exchange. Authorization refers to the logic granting permissions to access specific resources.

\noindent\textbf{Resource Identifiers} \\
Resource identifiers are a dedicated part of the message uniquely indicating which information the raw data should contain. The two main types are structured identifiers representing a relation (e.g.~web URLs) and independent identifiers (e.g.~UUID4). The former may be converted to independent identifiers using injective hash functions.

\noindent\textbf{Model Browsing} \\
If a protocol supports a browsable model, the underlying resource model and potential metadata can be retrieved over the wire in addition to the resource's data. Prominent examples are annotated REST APIs and OPC UA's browsing capabilities.

\noindent\textbf{CRUD Data Interaction}  \\
\textit{Create}, \textit{Read}, \textit{Update} and \textit{Delete} are the basic operations for interfacing from a data-oriented view. A protocol is classified as CRUD-interactive if the messages contain a part attributing the data access method. 

\noindent\textbf{Function Invocation}  \\
In addition to CRUD operations, a protocol may offer a method dedicated to function invocation on the message receiving party, e.g.~leading to physical state changes of a system or data processing before replying.

\begin{figure}[tb]
    \begin{center}
   \includegraphics[width=0.88\textwidth]{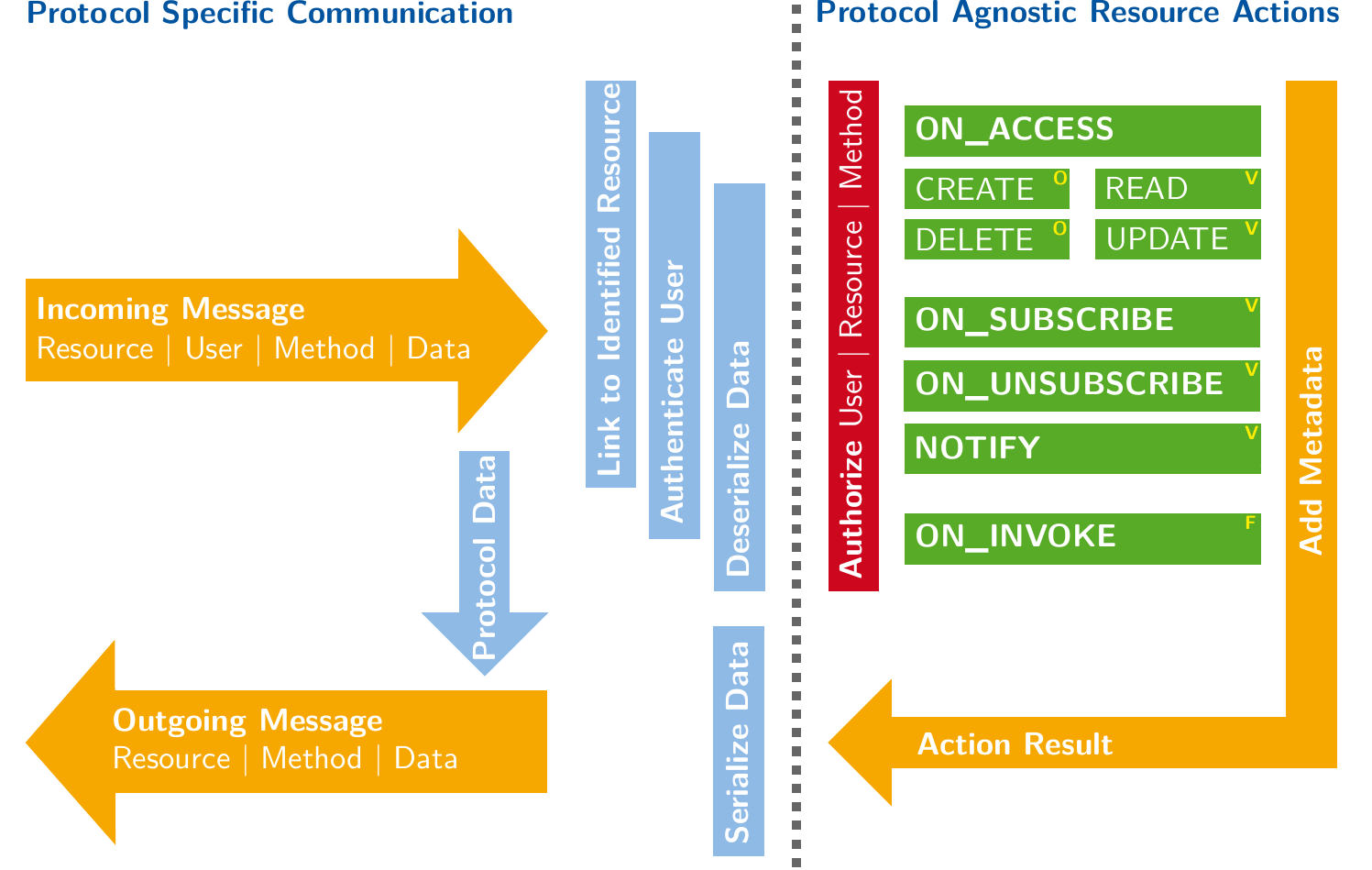}
    \end{center}
   \caption{Message exchange flow with resource actions: For an incoming message, resource identifier, user and method are extracted from the protocol specific part and the raw data is deserialized to primitive data fields. After an authorization based on the combination of resource identifier, user and method, the action linked to the method is called. Its result is enriched with metadata (covariance, timestamp, nonce) and serialized to the outgoing message. Protocol specific parts of the protocol may be generated and passed on to the outgoing without interaction with the resource (e.g.~session identifiers). A serialized representation of an internal error occurring during the action is a valid response.}
   \label{fig:message}
\end{figure}

Table \ref{tab:protocol_characteristics} assesses these properties for selected IoT protocols. Figure \ref{fig:message} outlines the message processing to a set of explicit actions required by the microservice resource implementation. They are utilized to establish an abstraction layer decoupling latter implementation from the use of a specific protocol. A mandatory prerequisite therefore is a model of the functions and data as resources of the microservice represented in serializable fields organized using unique (preferably structured) identifiers. 
Table \ref{tab:actions} explains this set of actions the resource implementation must react to. At the same time, the actions are matched with different protocol-specific methods. To reduce complexity, the authors propose a division of resources into \textit{objects}, \textit{functions} and \textit{variables} reacting to specific subsets: Objects provide an object-oriented resource model and are the only resources allowed to possess dynamic child resources. Functions are resources that are callable with a set of argument and return values. Variables encode read and potential write access to primitive data fields.
Authentication is regarded as protocol-specific task, while authorization is interpreted as resource-specific question if an action is allowed for a given combination of user and resource identifier and therefor not discussed in further detail.

Utilizing the abstraction layer, a device whose resources can be modeled in objects, functions and variables for which its internal software implements the respective subset of actions can be linked to any arbitrary protocol. The only requirements imposed are the expression of methods as aforementioned actions and handling of resource identification, authentication and serialization. Moreover, simultaneous interaction over multiple protocols is enabled as well as the deployment of adapters between protocols \cite{Pfrommer.2016, Evertz.2015}.

\section{FUNCTIONAL MODEL OF LARGE-SCALE METROLOGY INSTRUMENTS}
\label{sec:model}
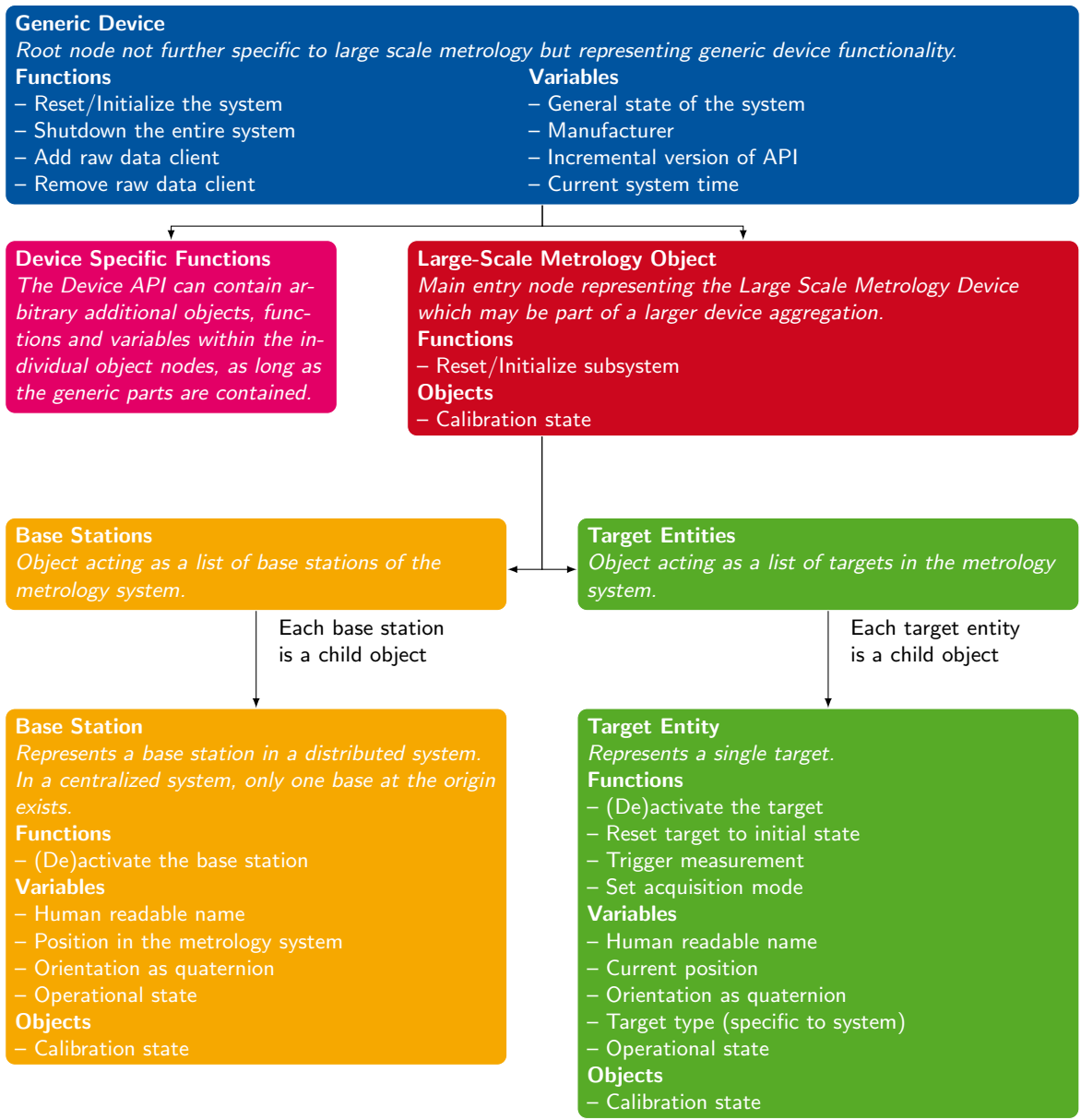
\begin{figure}[bt]
    \begin{center}
    \usetikzlibrary{positioning, shapes, calc}
    \tikzset{block/.style = {shape=rectangle, rounded corners,
                           draw,
                           align=left,
                           minimum height = 0.7cm,
                           top color = RWTHBlau,
                           bottom color = RWTHBlau,
                           color=white,
                           inner sep=4pt,
                           anchor = mid,
                           font = \small \sffamily}
    }
    \begin{tikzpicture}
    \node [block, text width=15.2cm] (device) {
        \textbf{Generic Device}
        
        {\footnotesize
        \textit{Root node not further specific to large scale metrology but representing generic device functionality.} \\
        \begin{minipage}[t]{0.48\textwidth}
        \textbf{Functions} \\
            -- Reset/Initialize the system \\
            -- Shutdown the entire system \\
            -- Add raw data client \\
            -- Remove raw data client
        \end{minipage}
        \begin{minipage}[t]{0.48\textwidth}
        \textbf{Variables} \\
        -- General state of the system \\
        -- Manufacturer \\
        -- Incremental version of API \\
        -- Current system time
        \end{minipage}
    }
    
    };
    \node [block, below right=0.5cm and 0cm of device.south west, text width=4.5cm, top color =RWTHMagenta, bottom color = RWTHMagenta]  (device_specific_functions){
        \textbf{Device Specific Functions}
    
        {\footnotesize \textit{The Device API can contain arbitrary additional objects, functions and variables within the individual object nodes, as long as the generic parts are contained.}}
    };
    \node [block, below right=0cm and 1cm of device_specific_functions.north east, text width=9.4cm, top color =RWTHRot, bottom color = RWTHRot]  (lsm_functions)  {
        \textbf{Large-Scale Metrology Object}
    
        {\footnotesize
            \textit{Main entry node representing the Large Scale Metrology Device which may be part of a larger device aggregation.}
            \\ \textbf{Functions} \\
                -- Reset/Initialize subsystem 
            \\ \textbf{Objects} \\
            -- Calibration state \\
        }
    
    };
    \node [block, below right=4.5cm and 0cm of device.south west, text width=6.95cm, top color =RWTHOrange, bottom color = RWTHOrange]  (lsm_bases)  {
        \textbf{Base Stations}
    
        {\footnotesize
        \textit{Object acting as a list of base stations of the metrology system.}
    }
    };
    \node[below right=0cm and 0.2cm of lsm_bases.south, font=\sffamily\footnotesize, text width=2.7cm] (lsm_bases_text) {Each base station is a child object};
 
    \node [block, below right=0cm and 1cm of lsm_bases.north east, text width=6.95cm, top color =RWTHGruen, bottom color = RWTHGruen]  (lsm_entities)  {
        \textbf{Target Entities}
        
            {\footnotesize
            \textit{Object acting as a list of targets in the metrology system.}
            }
 
    };
 
    \node[below right=0.0cm and 0.2cm of lsm_entities.south, font=\sffamily\footnotesize, text width=2.7cm] (lsm_entities_text) {Each target entity is a child object};
 
    \node [block, below right=1.4cm and 0cm of lsm_bases.south west, text width=6.95cm, top color =RWTHOrange, bottom color = RWTHOrange]  (lsm_base)  {
        \textbf{Base Station}
        
        {\footnotesize
        \textit{Represents a base station in a distributed system. In a centralized system, only one base at the origin exists.}
        \\ \textbf{Functions} \\
        -- (De)activate the base station \\
        \textbf{Variables} \\
        -- Human readable name \\
        -- Position in the metrology system \\
        -- Orientation as quaternion \\
        -- Operational state 
        \\ \textbf{Objects} \\
        -- Calibration state 
        }
        
        };
    \node [block, below right=0cm and 1cm of lsm_base.north east, text width=6.95cm, top color =RWTHGruen, bottom color = RWTHGruen]  (lsm_entity)  {    
        \textbf{Target Entity}
    
        {\footnotesize
        \textit{Represents a single target.}
        \\ \textbf{Functions} \\
        -- (De)activate the target \\
        -- Reset target to initial state \\
        -- Trigger measurement \\
        -- Set acquisition mode \\
        \textbf{Variables} \\
        -- Human readable name \\
        -- Current position \\
        -- Orientation as quaternion \\
        -- Target type (specific to system) \\
        -- Operational state
        \\ \textbf{Objects} \\
        -- Calibration state 
        }
    
        };
    
        \path[draw ,-latex] (device.south) |- ($(device_specific_functions.north) + (0cm, 0.2cm)$) -| (device_specific_functions.north);
        \path[draw ,-latex] (device.south) |- ($(lsm_functions.north) + (0cm, 0.2cm)$) -| (lsm_functions.north);
    
        \path[draw ,-latex] ($(lsm_functions.south) - (2.9cm, 0cm)$) |- ($(lsm_bases.north east) - (0cm, .75cm)$);
        \path[draw ,-latex] ($(lsm_functions.south) - (2.9cm, 0cm)$) |- ($(lsm_entities.north west) - (0cm, .75cm)$);
    
        \path[draw, -latex] (lsm_bases.south) -- (lsm_base.north);
        \path[draw, -latex] (lsm_entities.south) -- (lsm_entity.north);
    
    \end{tikzpicture}
    \end{center}
    \caption[Structural overview of unified interface]{Structural overview of the Large-Scale Metrology instrument interface model. The structure can be extended by specific implementations, as long as it contains the presented hierarchy to adhere to the proposed unification. Covariance, timestamp and a nonce are metadata that should be added to each measured variable. A detailed reference of the entity object is provided in appendix \ref{sec:appendix}.}
    \label{fig:interface_device_api_structure}
    \end{figure}

The interface implementation of a Large-Scale Metrology instrument can be decoupled from a specific protocol if its functionality is modeled in objects, functions and variables as stated in section \ref{sec:protocols}. Therefor two modeling perspectives exist: A physical view tightly linked to the device representing its physical layout and a functional view oriented along the information and functions (i.e.~resources) expected from the device. For the simple example of a traditional 6DOF robot these perspectives translate to either modeling its position command interface as six individually controllable joints or as kinematic capable of moving its tool center point in three directions and rotate around three axes. While the first view may result in a more accurate description , the latter perspective offers the advantage of uniformly modeling different devices providing the same capabilities in a black-box manner and thereby enabling their interchangeable use, e.g.~realizing a kinematic with six degrees of freedom by either a robot or a machine tool. 

Thinking of instruments as microservices, a functional view around the position information\footnote{\textit{Position information} is regarded as the measured position, orientation (if supported), associated covariance and, if applicable, traceability details.} of the system's mobile entities presents an essential part of a possible model-based interface. At the same time, \textsc{Franceschini} et al.~classify LSM instruments according to their hardware organization into \textit{distributed} and \textit{centralized} systems \cite{Franceschini.2014}. This offers a device-agnostic physical perspective for the interface if centralized systems are seen as systems with exactly one measuring station. Both approaches are combined in the interface model proposed in figure \ref{fig:interface_device_api_structure} using an object-oriented approach with base stations and target entities as main classes. The former are instantiated by the measuring stations of the systems, e.g.~indoor GPS transmitter, ultrawideband anchors or a laser tracker head.  

Instances of the latter are the mobile units whose position is actually determined by the instrument, e.g.~physical computing entities (PCEs) of an indoor GPS or spherical mounted retroreflectors (SMRs). These provide the generic, consistently identified interface to their current position information, regardless whether they are passive or active units\footnote{For example, the position of a SMR is measured by the angular encoders and the distance measurement system in the head of the laser tracker but is interfaced via the SMR's object instance.}.

An object is interpreted as class instantiation if it contains all functions, variables and child objects defined in the class. This allows to add device-specific (i.e.~inferred from a physical view) resources to the individual model-based interface while maintaining the generic functional perspective, e.g.~adding digital in- and outputs of indoor GPS PCEs as variables and effectively subclassing the entity class. Especially regrading the loosely defined \textit{Large Scale Metrology Object} and \textit{Generic Device Object} (cf.~figure \ref{fig:interface_device_api_structure}), adding additional device resources allows for seamless embedding of the proposed interface into more complex device models.

Probes are explicitly not covered by the proposed model as they are interpreted as devices themselves which rely on one or more target entities at defined physical interfaces. Moreover, a single instrument may be used with different types of probes potentially even unknown to the manufacturer. Following the microservice paradigm, a probing coordinate measurement system could be realised by a superordinate software accessing the position information of the Large-Scale Metrology instrument over its interface after a trigger signal has been received from the probe's device interface.

The yet not further defined calibration objects account for traceable measurements on a physical and cybersecure level. It is intended to digitally represent a calibration certificate including a cryptographinc certificate in primitive fields, which is currently subject to research \cite{Eichstadt.2018}.

\section{VALIDATION FOR LASER TRACKER}
\label{sec:tracker}
\begin{figure}[ht]
    \includegraphics[width=\textwidth]{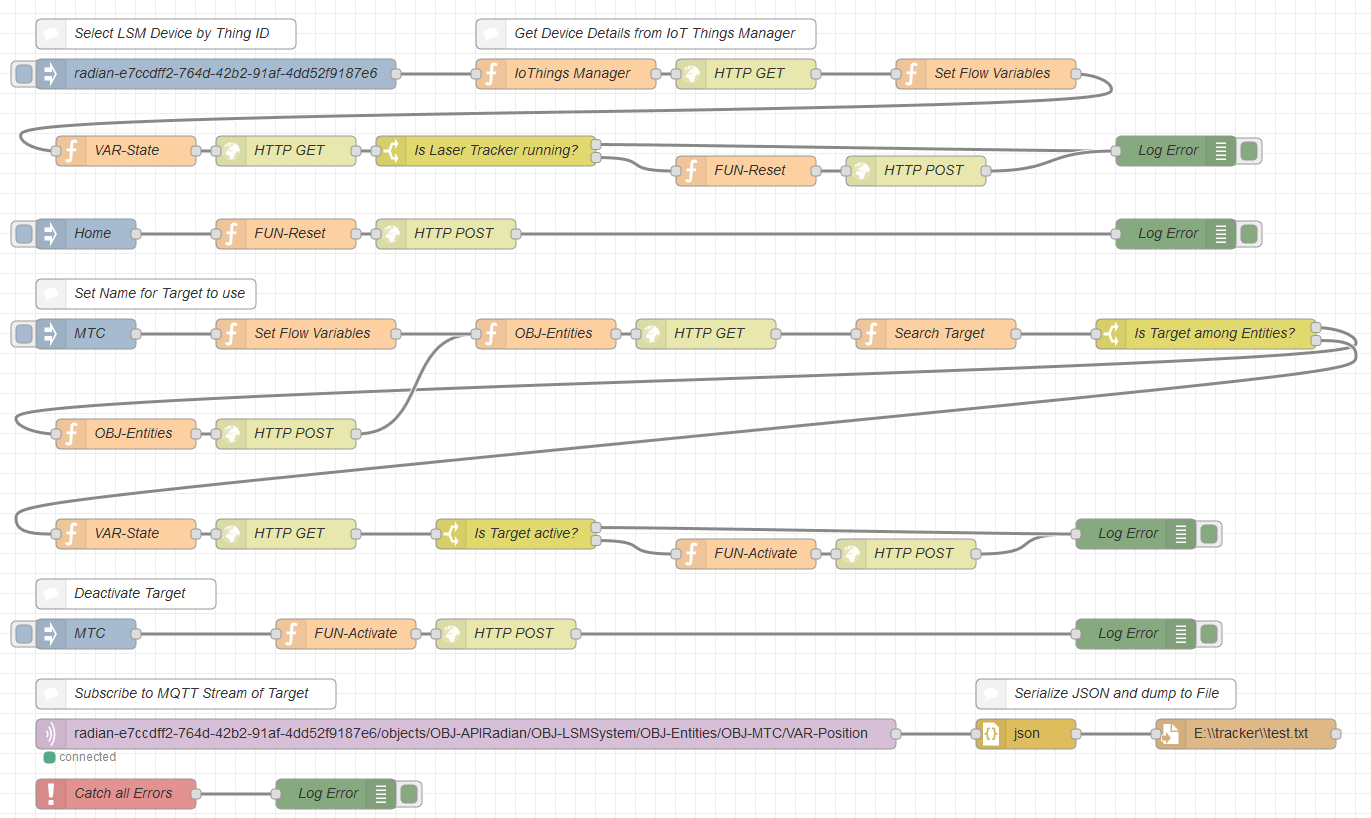}
    \label{fig:nodered}
    \caption{Example flow created with Node-RED using an API Radian\textsuperscript{TM} laser tracker for Machine Tool Calibration via a HTTP/REST and MQTT interface. The information necessary to connect to the device (e.g.~network address) is obtained from a IoT managing service.}
    \end{figure}
The model-based interface approach described in the previous section was evaluated for an API Radian\textsuperscript{TM} laser tracker. Its head is interpreted as the only base station defining the local coordinate system. The mobile entities consist of SMRs and other compatible targets, e.g.~smart targets also measuring orientation. Multiple targets are used for convenience to facilitate automated network measurements and maintain the analogy to multi-target systems. 

\noindent Due to the measurement principle, only one entity can be in active operational state at the time. Activating an SMR corresponds to pointing the laser tracker to its last known position and let it run its internal search routine. The operational state of the target itself corresponds to whether the laser tracker is logged in and able to measure. The initial/reset state of the instrument represents a warmed-up laser tracker logged into its home SMR. Three acquisition modes are supported: \textit{Continuous} means that measurements are taken and dispatched to the abstraction layer (for read and notify actions) continuously and as fast as possible. \textit{Triggered} and \textit{External} acquisition modes only dispatch the measurement after a software respectively hardware trigger signal. Covariances are estimated using a model of the device based on the raw data obtained for the distance measurement and the angular encoders. Resetting a target re-runs the internal search routine.
Additional functions such as a jogging the tracker or toggling the built-in camera are implemented by extending the base station object.

Figure \ref{fig:nodered} shows an flowchart for using a Large-Scale Metrology instrument's continuous data stream taken from the application of calibrating a machine tool which was applied using a laser tracker  interfacing the abstraction layer in a cloud-based Node-RED instance via HTTP/REST and MQTT \cite{Montavon.2019, Blackstock.2014}. In a second use-case, an OPC UA based interface was used to integrate the same laser tracker into a robotic research platform to measure the deformation of a CFRP airplane panel \cite{Schmitt.2014}. Both use-cases showed that the device model introduced significantly reduced integration times. For future applications, two laser trackes have been added as a permanently available services to the network infrastructure of the authors' laboratory shop floor.
\section{CONCLUSIONS AND OUTLOOK}
The proposed interface model for Large-Scale Metrology systems allows to decouple the implementation and structure of the instrument's resources from the use of a specific communication protocol. Subseqently, applications can be designed independent of the used measurement device. In combination, this facilitates their integration into complex, heterogeneous industrial systems. Leveraging the relatively small set of requirements to the instrument's minimal implementation of the defined classes, the authors will apply the interface model to a Nikon iGPS\textsuperscript{TM} and Pozyx\textsuperscript{TM} ultrawideband systeme, such that these are available as sensor microservice to cyberphysical applications.

\acknowledgments 
The authors acknowledge funding from the LaVA project (Large Volume Applications, contract 17IND03 of the European Metrology Programme for Innovation and Research – EMPIR). The EMPIR initiative is co-funded by the European Union's Horizon 2020 research and innovation programme and the EMPIR Participating States.
The authors would like to thank the German Research Foundation DFG for the kind support within the Cluster of Excellence \textit{Internet of Production} - Project-ID: 390621612.

\bibliography{references} 
\bibliographystyle{spiebib} 

\clearpage
\appendix
\section{Entity Instrument Model Details}
\label{sec:appendix}
\begin{table}[h]
    \renewcommand{\arraystretch}{1.25}
    \renewcommand{\arrayrulewidth}{1.25pt}
    \arrayrulecolor{white}
    \begin{tabular}{|>{\columncolor{RWTHCustomGrau}}p{0.97\textwidth}|} \hline \rowcolor{RWTHHellblau}
        \textbf{FUN-Activate (bool \textit{active} $\to \emptyset$) }   \\ \hline
        Activate or deactive the entity according to \textit{active}. Activating an entity should put the entity into a state that measurements are enabled. \\ \hline
        \textbf{Laser Tracker:} Point to the last position known for the reflector and if available, start target search. \\ \hline
    \end{tabular}

    \begin{tabular}{|>{\columncolor{RWTHCustomGrau}}p{0.97\textwidth}|} \hline \rowcolor{RWTHHellblau}
        \textbf{FUN-Reset ($\emptyset \to \emptyset$) }   \\ \hline
        Reset the entity to its initial state. \\ \hline
        \textbf{Laser Tracker:} Restore the initial position of the reflector. \\ \hline
    \end{tabular}

    \begin{tabular}{|>{\columncolor{RWTHCustomGrau}}p{0.97\textwidth}|} \hline \rowcolor{RWTHHellblau}
        \textbf{FUN-Trigger (int \textit{count}, string \textit{nonce} $\to \emptyset$) }   \\ \hline
        Trigger the entity to measure \textit{count} times and set the internal nonce to \textit{nonce}. This call is only valid for an active target in triggered acquisition mode. \\ \hline
        \textbf{Laser Tracker:} Dispatch the next \textit{count} valid measurement results and associate these with target in question.   \\ \hline
    \end{tabular}

    \begin{tabular}{|>{\columncolor{RWTHCustomGrau}}p{0.97\textwidth}|} \hline \rowcolor{RWTHHellblau}
        \textbf{FUN-Acquisition (enum \textit{mode}, string \textit{nonce} $\to \emptyset$) }   \\ \hline
        Set the acquisition mode to \textit{mode} $\in$ [CONTINUOUS, TRIGGERED, EXTERNAL]. Continuous means the measurements are taken continuously. In triggered mode, measurements are triggered either by software or an external signal. If the instrument does not physically support triggering, the software should wait for the next value.\\ \hline
        \textbf{Laser Tracker:} Trigger readout of distance measurement and rotary encoders and save the result for the activated target.  \\ \hline
    \end{tabular}

    \begin{tabular}{|>{\columncolor{RWTHCustomGrau}}p{0.97\textwidth}|} \hline \rowcolor{RWTHHellblau}
        \textbf{VAR-Name (string) }   \\ \hline
        Human readable name of the target. \\ \hline
    \end{tabular}

    \begin{tabular}{|>{\columncolor{RWTHCustomGrau}}p{0.97\textwidth}|} \hline \rowcolor{RWTHHellblau}
        \textbf{VAR-Position (double[3]) }   \\ \hline
        Last measured position of the target in metres. \\ \hline
    \end{tabular}

    \begin{tabular}{|>{\columncolor{RWTHCustomGrau}}p{0.97\textwidth}|} \hline \rowcolor{RWTHHellblau}
        \textbf{VAR-Quaternion (double[4]) }   \\ \hline
        Last measured orientation of the target as quaternion. The value should be set to (1,0,0,0) if it is not measured.\\ \hline
    \end{tabular}

    %

    \begin{tabular}{|>{\columncolor{RWTHCustomGrau}}p{0.97\textwidth}|} \hline \rowcolor{RWTHHellblau}
        \textbf{VAR-State (enum) }   \\ \hline
        Current state of the target as choice of [TRIGGERED, CONTINUOUS, INACTIVE, EXTERNAL, WARNING, ERROR, MAINTENANCE].\\ \hline
    \end{tabular}
    
    \begin{tabular}{|>{\columncolor{RWTHCustomGrau}}p{0.97\textwidth}|} \hline \rowcolor{RWTHHellblau}
        \textbf{OBJ-Calibration}   \\ \hline
        Calibration object yet not further defined\\ \hline
    \end{tabular}

    \caption{Description of the variables (VAR-), functions (FUN-) and objects (OBJ-) of the entity class with example function mappings a laser tracker. The model of a base station is similar.}

\end{table}

\clearpage

\end{document}